\documentstyle[12pt]{article}
\advance\textheight by \topskip
\def\beq{\begin{equation}}
\def\eeq{\end{equation}}
\def\beqa{\begin{eqnarray}}
\def\eeqa{\end{eqnarray}}
\def\beq{\begin{equation}}
\def\eeq{\end{equation}}
\def\beqa{\begin{eqnarray}}
\def\eeqa{\end{eqnarray}}
\newcommand{\NP}[3]{{ Nucl. Phys.} {\bf B #1} {(19#2)} {#3}}

\newcommand{\be}{\begin{equation}}
\newcommand{\eq}{\end{equation}}

\begin{document}
\begin{flushright}
\hfill HUB-EP-97/18\\
\hfill hep-th/9703101\\
\end{flushright}
\vspace{1cm}
\begin{center}
\baselineskip=16pt
{\Large\bf GENERAL STATIC $N=2$ BLACK HOLES} 
\vskip 2cm
{\bf W. A. Sabra}\\
\vskip 0.5cm
\hfill

{\em Humboldt-Universit\"at zu Berlin,\\
Institut f\"ur Physik,\\ 
D-10115 Berlin, Germany}\\
E-mail:sabra@qft2.physik.hu-berlin.de
\vskip 0.2cm
\end{center}
\vskip 1 cm

\centerline{\bf ABSTRACT}
\begin{quotation}
We find general static BPS black hole solutions for general $N=2$, $d=4$ 
supergravity theories with an arbitrary number of vector multiplets. 
These solutions are completely specified by the K\"ahler potential of the 
underlying special K\"ahler geometry and a set of constrained harmonic 
functions.
\end{quotation}
\newpage
\baselineskip=15pt
Recently lots of activity \cite{fks, s, fk, ksw, group} 
have been devoted to the study of BPS black holes in ungauged $N=2$ 
supergravity models coupled to vector 
multiplets\footnote{For a review on $N=2$ supergravity and special geometry
see for example \cite{freandvan}}. The underlying special 
geometry structure of the $N=2$ theory has played an essential role 
in the analysis of the Bekenstein-Hawking entropy \cite{bh} of the $N=2$ 
black hole solutions. 
The special geometry of the couplings of $n$ vector multiplets in 
$N=2$ supergravity is completely determined by the covariantly holomorphic 
symplectic sections $(L^I, M_I),$ $(I=0, \cdots, n)$,  satisfying the 
symplectic constraint
\be
i(\bar L^IM_I-L^I\bar M_I)=1
\eq 
and depend on the complex scalar fields of the
vector multiplets $z^{i}$ ($i=1,\cdots,n$). These scalars 
parametrise a special
K\"ahler manifold with the metric $g_{i j^\star}=\partial_{i}
\partial_{j^\star} K(z,\bar{z})$, where $K(z,\bar{z})$ is the K\"ahler
potential expressed by
\be 
e^{-K} =
i (\bar{X}^I F_I - X^I \bar{F}_{I})
\eq
and  $(X^I, F_I)$ are holomorphic sections defined by
\be
X^I=e^{-{K\over2}}L^I,\quad F_I=e^{-{K\over2}} M_I.
\eq 
The relation between $F_I$ and $X^I$ depends very much on the particular 
embedding of the isometry group of the underlying special K\"ahler manifold 
into the symplectic group \cite{sym}. 
In some cases, $F_I$ can be expressed as 
a first derivative (with respect to $X^I$) of a holomorphic prepotential.

In general $N=2$ supersymmetric 
theories, the mass of a BPS state 
which breaks half of the supersymmetry,
can be expressed in terms of the sections 
by 
\be
M^2_{BPS}=|Z|^2 = e^K|q_I X^I - p^I F_I|^2=|q_I L^I - p^I M_I|^2
\eq
where $q_I$ and $p^I$ are electric and magnetic charges 
corresponding to the gauge group 
$U(1)^{n+1},$ with $n$ the number of vector multiplets and the extra 
$U(1)$ factor is due to the graviphoton, 
and $Z$ is the central charge of the underlying $N=2$ supersymmetry algebra.   

The black hole ADM mass, which in the $N=2$ theory depend on the electric 
and magnetic charges as well as the value of the scalar fields at spatial
infinity, is given by 
\be
M^2_{ADM}=\vert Z_\infty\vert^2=\vert Z(z_\infty^A, \bar z_\infty^A)\vert^2.
\eq
Here $z^i_\infty$ $(i=1,\cdots,n),$ are the values of 
the moduli at spatial infinity.

A fundamental result has been obtained \cite{fk} which provides
an algorithm for the macroscopic determination of Bekenstein-Hawking entropy
for extremal $N=2$ black holes. One simply extremize the central charge at 
fixed charges, the extremum value then gives the Bekenstein-Hawking 
entropy $S$ via the relation
\be
S=\pi\vert Z_{fix}\vert^2.
\eq 

Moreover, it was found that the horizon acts as an attractor for the 
scalar fields, which means
that the values of the moduli are fixed at the horizon and are 
independent of their values at spatial infinity.
For purely magnetic case, the values where shown to depend on ratios of 
the charges \cite{fks}. This was later shown to be a general phenomenon 
which extends to dyonic solutions \cite{s,fk}. 
In general the fixed values of the scalars are those which
extremize the central charge. Therefore, $N=2$ black holes lose all 
their scalar hair near the horizon and are uniquely characterised by conserved 
charges corresponding to the gauge symmetries. For N=2 supergravity models
obtained from Calabi-Yau three-folds compactification, the holomorphic
prepotential and hence the entropy depends on the 
the classical intersection 
numbers, Euler number and rational instanton numbers of the underlying  
Calabi-Yau three-fold.

The above results come about by using the supersymmetry transformations laws
for the gravitino and gauginos in ungauged bosonic part of $N=2$ supergravity. 
Demanding unbroken supersymmetry \cite{fk} near the horizon, 
gives the result that the covariant derivative of the central charge has
to vanish. At this point, the moduli become functions of the charges only. 
The resulting equations which define the moduli in terms of the charges
are given by
\be
i(\bar ZL^\Sigma-Z\bar L^\Sigma)=p^\Sigma,\qquad 
i(\bar ZM_\Sigma-Z\bar M_\Sigma)=q_\Sigma.
\eq
These equations can also be obtained by solving the equations of motion for
the double-extreme black hole solutions, where the scalars are assumed to be
constants everywhere \cite{ksw}.

While supersymmetric solutions of pure $N=2$ supergravity are known \cite{tod},
supersymmetric solutions in $N=2$ supergravity coupled to vector multiplets 
and hypermultiplets are not fully understood. For the latter theories, 
only simplified purely magnetic black hole solutions have been constructed
\cite{fks}. However, in \cite{fks} it 
was realized that the axion-dilaton black hole solutions constructed in
\cite{ko} can be reinterpreted as solutions of $N=2$ 
supergravity coupled to one vector multiplet, 
namely the $SU(1,1)\over U(1)$ model. One simply
sets the holomorphic coordinates to complex harmonic 
functions and then the metric solution
can be expressed in terms of the K\"ahler potential. Later it was 
realised in \cite{bro} that the 
conditions imposed on the harmonic functions for the axion-dilaton black hole
solution in \cite{ko} correspond in the $N=2$ language to the vanishing of 
the K\"ahler connection. A non-vanishing K\"ahler connection leads
to the supersymmetric Israel-Wilson-Preje\'s like solution \cite{iwp}
for the dilaton-axion supergravity \cite{bro}. Moreover in \cite{bss} 
black hole solutions were found to be expressed in terms of harmonic 
functions. 

All the above observations provide an insight for the study of static 
black hole and stationary 
solutions for the more general $N=2$ supergravity theory with arbitrary 
number of vector multiplets. In this work and as a starting point in 
this direction, we will
focus on the study of static black hole solutions and report on the 
stationary solutions in a future publication.
Our analysis is completely independent of the existence of
holomorphic prepotential and  
formulated entirely in terms of the holomorphic sections $(X^I, F_I)$. 
The main result is that the K\"ahler potential provides the 
expression for the metric and that the imaginary part of the holomorphic
sections are given by a set of harmonic functions which depend on the 
electric and magnetic charges of the model. Details of this work will be 
presented elsewhere.

We will ignore the hypermultiplets and assume that they are set to 
constant values. In this case, 
the bosonic $N=2$ supergravity action includes one gravitational 
and $n$ vector multiplets and 
is given by
\be
{\cal L}_{N=2} =\int \sqrt{-g}\  d^4x\Big(-{1\over2}R+g_{i {{j}^\star}}
\nabla^{\mu} z^i \nabla _{\mu} \bar z^{{j}^\star}  
+ i\left(\bar {\cal N}_{\Lambda \Sigma} 
{\cal F}^{- \Lambda}_{\mu \nu}
{\cal F}^{- \Sigma \vert {\mu \nu}}
\, - \,
{\cal N}_{\Lambda \Sigma} {\cal F}^{+ \Lambda}_{\mu \nu}
{\cal F}^{+ \Sigma \vert {\mu \nu}} \right )  
\label{ni}
\eq
where 
${\cal N}_{\Lambda \Sigma}$ is the symmetric period matrix whose imaginary
and real part are, respectively, related to the Lagrangian 
kinetic and topological terms of the vector fields, and 
${\cal F}^{+ \Lambda}_{\mu \nu}$, ${\cal F}^{- \Lambda}_{\mu \nu}$ are 
respectively, the self-dual and anti-self-dual vector field strengths.

It is known \cite{tod} in general that for $N=2$ theories, 
a static metric
admitting supersymmetries can be put in the Majumdar-Papapetrou metric
form \cite{mp}, 
\be
ds^2 =e^{2 U} dt^2-e^{-2U} d\vec{x} d\vec{x}, \qquad\eq
Here we consider spherically symmetric solutions, $i.e,$ $U$ is only a 
function of $r\equiv \sqrt {\vec{x}.\vec{x}}.$

To find the explicit BPS black hole solution 
it is more convenient to solve the 
supersymmetry transformations since these transformation rules
depend linearly on the first derivatives of the bosonic fields and thus
provide first order differential equations.
The supersymmetry transformation rules
for the gravitino and gauginos in a bosonic background of 
$N=2$ supergravity
are given by \cite{freandvan}
\begin{eqnarray}
\delta\,\psi_{\alpha\mu} &=& \nabla_\mu \epsilon_\alpha -
\frac{1}{4} T^-_{\rho\sigma} \gamma^{\rho\sigma}
\, \gamma _\mu \, \varepsilon_{\alpha\beta}\epsilon^\beta, 
\\
\delta\lambda^{i\alpha} & = & 
i\nabla_\mu z^i\gamma^\mu\epsilon^\alpha 
+ G^{-i}_{\rho\sigma}\gamma^{\rho\sigma}
\varepsilon^{\alpha\beta}\epsilon_\beta
\end{eqnarray}
where $\psi_{\alpha\mu}$ and $\lambda^{i\alpha}$ are the chiral gravitino 
and gauginos fields, $\epsilon_\beta,$ $\epsilon^\beta$ are the chiral and
antichiral supersymmetry parameters respectively and 
$\varepsilon ^{\alpha\beta}$ is the $SO(2)$ Ricci tensor.
The quantities $T^-_{\rho\sigma}$ and $G^{-i}_{\rho\sigma}$ are the field
strengths of the gravi-photon and matter-photon, respectively.

>From the conditions of vanishing supersymmetry transformation, $i.e.,$
$\delta\psi _{\alpha\mu} =\delta\lambda^{i\alpha}= 0,$ one obtain 
for a particular choice of the supersymmetry parameter, the following
two equations \footnote{These equations were also
derived very recently in \cite{fkg} using a different approach.}\cite{f} 
\begin{eqnarray}
{dU\over dr}& =& {Ze^{U}\over r^2},\\
\label{susan}(\partial_i+{1\over 2}\partial_i K)L^\Sigma{dz^i\over dr}&=& 
-{e^{U}\over r^2}\Big(Z\bar L^\Sigma-{1\over2}(ip^\Sigma-
(Im {\cal N}^{-1})^{\Sigma\Lambda}q_\Sigma+
(Im {\cal N}^{-1})^{\Sigma\Gamma} Re 
{\cal N}_{\Gamma\Delta}p^\Delta)\Big)
\nonumber
\end{eqnarray}

The solution of the above equations of course depends on the particular model
one is considering, $i.e.,$ the choice of the special K\"ahler 
manifold. In what follows we will solve the above
differential equations in a model independent way. 
Our ansatz for the solution is to take
\be
e^{-2U}=e^{-K}=i(\bar X^IF_I-X^I{\bar F}_I)
\label{sr}
\eq
Then the first differential equation in (12) gives 
\be
{d\over dr}e^{-2U}=-2{(X^Iq_I-F_Ip^I)\over r^2}
\label{r}
\eq

If we further demand that our solution must satisfy the following relation
\be
\bar X^I{dF_I\over dr}-{dX^I\over dr}\bar F_I=
{d\bar X^I\over dr}F_I-{X^I}{d\bar F_I\over dr},
\label{vcc}
\eq
then from (\ref{sr}) we obtain
\be
{d\over dr}e^{-2U}=2i({d\bar X_I\over dr}F_I-X^I{d\bar F_I\over dr}).
\label{vc}
\eq
If we write
\be
i(X^I-\bar X^I)=f^I, \qquad i(F_I-\bar F_I)=g_I
\label{es}
\eq
where $f^I$ and $g_I$ are real functions  which only depend on 
the radial distance $r$, eq. (\ref{vc}) can then be rewritten as
\be
{d\over dr}e^{-2U}=2\Big(X^I{dg_I\over dr}-F_I
{df^I\over dr}\Big)
\label{s}
\eq
where we made use of the relation 
\be
{dX^I\over dr}F_I-X^I{dF_I\over dr}=0.\eq
which follows from the underlying special 
geometry structure.

Comparing the equation (\ref{s}) with (\ref{r}), one immediately 
arrive at the following solution
\be
f^I={\tilde h^I}+{p^I\over r}, \qquad g_I={h_I}+{q_I\over r}
\eq
where $h_I$ and ${\tilde h}^I$ are constants related to the values of the 
holomorphic sections (scalars) at infinity. These constants are restricted by
demanding that the metric is asymptotically flat. One can also demonstrate 
that our solution solves the second differential equation in (12). 

In deriving our solutions, we had to impose the condition (\ref {vcc}), which
is nothing but the condition of the vanishing of the 
the K\"ahler connection which can be expressed in the following form
\be
A_\mu={-i\over2}\Big(\partial_i K\partial_\mu z^i-
\partial_{i^\star} K\partial_\mu {\bar z}^{i^\star}\Big).
\eq
Using (\ref{es}), the vanishing of $A_\mu$ corresponds to
\be
f^I{dg_I\over dr}-g_I{df_I\over dr}=0.
\eq
This non-trivial condition does not impose any constraints on the electric and
magnetic charges and only puts an additional constraint on $h_I$ and 
${\tilde h}^I$, and 
ensures that the black hole saturates the BPS bound. 

In conclusion, we have found general supersymmetric static black hole solutions
for $N=2$ supergravity theories coupled to an arbitrary number of 
vector multiplets. These solutions 
are completely determined in terms of the underlying K\"ahler geometry
of the moduli space of the scalars of the vector-multiplets. The  metric is 
expressed in terms of the K\"ahler potential of the theory where the imaginary
part of the holomorphic sections are given by harmonic functions. In summary

\begin{eqnarray}
ds^2 &=&e^{2 U} dt^2 -  e^{-2 U} d\vec{x} d\vec{x},\\ 
\nonumber
e^{-2U}&=&e^{-K}=i(\bar X^IF_I-X^I\bar F_I)\\
\nonumber
Im\pmatrix{X^I\cr F_I}&=&-{1\over2}\pmatrix{{\tilde h}^I+{p^I\over r}\cr 
{h}_I+{q_I\over r}}\\
\nonumber
{\tilde h}^Iq_I-{h}_Ip^I&=&0,\qquad e^{-K_\infty}=1.
\end{eqnarray}
We believe that our results are essential to the study of 
perturbative and
nonperturbative corrections for $N=2$ black hole 
solutions as well as the 
study of massless black holes. 

\section*{Acknowledgements}
This work is supported by DFG and in part by DESY-Zeuthen. I would like to
thank K. Behrndt and T. Mohaupt for useful discussions.

\end{document}